\documentclass[conference]{IEEEtran}
\IEEEoverridecommandlockouts

\usepackage{cite}
\usepackage{amsmath,amssymb,amsfonts}
\usepackage[ruled,vlined]{algorithm2e}
\usepackage{graphicx}
\usepackage{textcomp}
\usepackage{xcolor}
\usepackage{tikz}
\usepackage{placeins}
\usetikzlibrary{arrows.meta, positioning, calc}

\def\BibTeX{{\rm B\kern-.05em{\sc i\kern-.025em b}\kern-.08em T\kern-.1667em\lower.7ex\hbox{E}\kern-.125emX}}

\usepackage{hyperref}
\hypersetup{
    colorlinks = true,
    citecolor = magenta,
    linkcolor = purple
}
\usepackage{xcolor}
\usepackage{tabularx, makecell, linegoal}

    \usepackage{ifthen}
    
\newboolean{ShowComments}
\setboolean{ShowComments}{true}  
\ifthenelse{\boolean{ShowComments}}%
	{
		\newcommand{\ColorComment}[3]{%
				{\colorbox{#1}{\color{white}   \textsf{\textbf{#2}}} \textcolor{#1}{#3}}}

	}%
	{
		\newcommand{\ColorComment}[3]{}

	}%

\definecolor{commentcolor}{RGB}{80, 180, 180}

\definecolor{amincolor}{RGB}{100, 120, 240}
\definecolor{amancolor}{RGB}{240, 120, 100}

\begin{document}
\title{A Distributed Switching Protocol \\ for Quantum Networks \\
{

} 

\thanks{This work was supported by JST [Moonshot R\&D Program] Grant Number [JPMJMS256K] and Grant Number [JPMJMS226C]}
}
\author{
\IEEEauthorblockN{
    Aman Yacob Tekleab\IEEEauthorrefmark{1},
    Yifeng Shen\IEEEauthorrefmark{2},
    Yoshii Yutaro\IEEEauthorrefmark{2},
    Amin Taherkhani\IEEEauthorrefmark{4},
    Rodney Van Meter\IEEEauthorrefmark{2}\IEEEauthorrefmark{3},
    Shota Nagayama\IEEEauthorrefmark{4}
}\\

\IEEEauthorblockA{\IEEEauthorrefmark{1}\textit{Faculty of Policy Management, Keio University Shonan Fujisawa Campus, Kanagawa, Japan}}

\IEEEauthorblockA{\IEEEauthorrefmark{4}\textit{Graduate School of Media Design, Keio University Hiyoshi Campus, Kanagawa, Japan}}
\IEEEauthorblockA{\IEEEauthorrefmark{3}\textit{Quantum Computing Center, Keio University, Kanagawa, Japan}}
\IEEEauthorblockA{\IEEEauthorrefmark{2}\textit{Faculty of Environment and Information Studies, Keio University Shonan Fujisawa Campus, Kanagawa, Japan}}

\{amanyacob, tomshen, y.yoshii\}@keio.jp, \{amin, rdv\}@sfc.wide.ad.jp, shota@qitf.org}

\maketitle
\begin{abstract}
With the advent of the construction and deployment of entanglement-based quantum networks, the efficient use of network resources will become a critical challenge for the scalable operation of such a system. Recently, architectures that incorporate memoryless optical switches have gained attention for forwarding entangled photons. By leveraging these architectures, costly resources such as high efficiency Bell State Analyzers (BSAs) can be shared across the network. Nevertheless, the introduction of switching substantially complicates the process of multiplexing and resource allocation compared to an individual link. In this work, we propose a switching protocol for unbuffered, multidrop quantum networks in a photonic synchronization domain that establishes a link between two end nodes using a shared BSA in the switched network. To achieve this, two end nodes cooperatively select the target BSA node with the lowest path cost and independently reserve each path within the network. Bi-path reservations are performed to allocate resources in a distributed manner. The proposed protocol is evaluated through simulation on Q-Fly network topologies under varying traffic conditions. The results demonstrate high link establishment success with stable performance even under increased network load. These capabilities which are driven by our proposed protocol are an essential way to realize large-scale, managed, and automated quantum networks.


\end{abstract}


\section{Introduction} \label{Intro}
 
Entanglement distribution is considered a primary function of quantum networks~\cite{vanmeter2014quantum}, supporting different applications such as distributed quantum computation, privacy-preserving delegated quantum computation, and secure communications~\cite{knorzer2025distributed,fitzsimons2017private,singh2023towards}. 

Although quantum repeaters and routers with quantum memory and the corresponding network-layer routing protocols \cite{van2013path, abane2025entanglement,lucho2011multiplexing,cicconetti2021request} are essential for long-distance wide-area networking, for medium scale and local-area quantum networks, efficient and flexible resource sharing between quantum nodes using reconfigurable optical interconnects such as optical switches is also important. This can be achieved through link architectures that include supporting nodes such as Bell state analyzers (BSAs) and/or entangled photon pair sources (EPPSs) without requiring stationary memory~\cite{jones2016design, beukers2024remote, soon2024implementation,main2025distributed}. This provides scalable communication among end nodes using existing high demand resources such as high efficiency Superconducting Nanowire Single Photon Detectors (SNSPDs) in BSAs. 

Moreover, recently emerging network testbeds have provided initial prototypes of quantum entangled networks at the scale of a few end nodes, with the ability to forward entangled photons using optical switches~\cite{alshowkan2021reconfigurable, monga2023quant, taherkhani2024scalable}. As the number of nodes and the switch radix increase, each node requires well-defined coordination to establish complete paths for entangled photons. Developing and extending these types of networks to multiple areas with hundreds of nodes requires appropriate resource allocation protocols, such as the switching protocol.

Depending on the level of control, a switching protocol can be divided into centralized and distributed approaches. In the former, a centralized controller handles incoming requests. That controller receives, manages, and schedules all requests to utilize the resources in the network. However, in the distributed approach, there are no centralized controllers. Therefore, all nodes are required to collaborate to handle demands in the network. \\

\subsection{Motivation}
\begin{figure}[htb]
    \centering
    \includegraphics[width=1\linewidth]{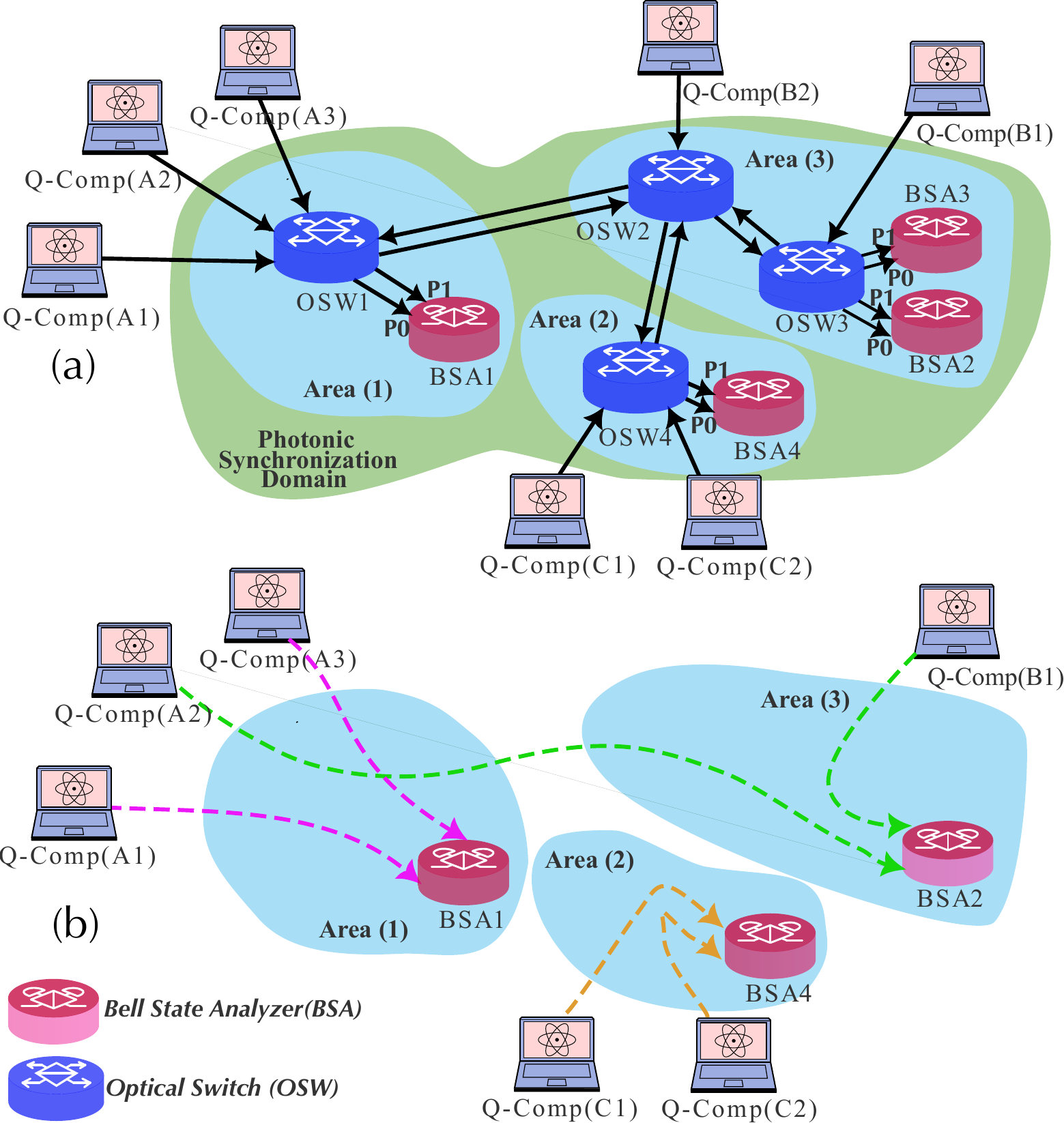}
    \caption{ 
    Establishment of links between quantum end nodes entities in an unbufferred multidrop switching network. (a) network of end nodes (quantum processing or sensing or measurement units) or middle nodes (quantum repeaters or routers) are connected via memoryless optical switches in one or multiple areas in a photonic synchronization domain. (b) Allocation of paths and BSAs in response to three requests for link establishment.   
    }
    \label{fig:sample-switching-in-QLAN}
\end{figure}

Recently, new architectures, routing protocols, and resource assignment schemes have been proposed for BSA pool allocation~\cite{koyama2024optimal, gauthier2025demand} or EPPS pool allocation~\cite{drost2016switching} through a single switch with centralized control. Those can extend the idea of quantum link establishment with two end nodes and one support node~\cite{dahlberg2019link}. However, forwarding of photons in a switched network composed of a chain of multiple optical switches across single or multiple areas has not been thoroughly investigated. The distributed form of switching comes from the need to scale to connecting tens to thousands of nodes in a multi-area network.  An example of such a network with an arbitrary topology is shown in Fig. \ref{fig:sample-switching-in-QLAN}. At the edge of the network, quantum nodes with stationary memory can be deployed, whereas the interconnection network relies on support nodes such as Bell state analyzers (or entanglement photon pair sources) for entanglement generation or interference~\cite{vanmeter2022quantum}. This region is referred to as a Photonic Synchronization Domain (PSD)~\cite{mori2024scalable}, where shared resources are configured to synchronize the arrival (or emission) of entangled photons so that the photons interfere, resulting in entanglement distribution between the end nodes.

The PSD can be deployed across multiple areas or in a single area to support quantum local area networks (QLANs). Each area can be understood as a physical zone on the scale of a campus or building which can operate under a single administrative domain. In the mid-term, a limited number of switching layers is expected to enable connectivity among tens to hundreds of nodes. Although such architectures may eventually scale to large, structured switching networks with thousands of nodes to shape the datacenter scale quantum networks~\cite{sakuma2024optical, shapourian2025quantum}, in this work we assumed that each pair of end nodes can run their applications independently of other remaining nodes. In this scenario, each node can initiate any entanglement request with any other nodes in the network at any time. 

\subsection{Multiplexing and Resource Allocation}
In parallel with photon forwarding, shared resources (support nodes such as BSAs and the shared links) are allocated based on incoming entanglement requests and local schedules of each resource. In such a network, the addition of switching complicates the process of multiplexing and resource allocation compared to an individual link or network of links separated by quantum memories with flexible timing, which decouples optical domains~\cite{lucho2011multiplexing}. 

The complexity of the problem also arises because  the forwarding path of each candidate affects the generation time of entanglements due to its associated cost. For example, in a time-division multiplexing scenario, if two nodes are connected to a BSA via switched circuits, the required allocation time for a request is significantly longer than in the case where the nodes are directly connected to the BSA. This is due to the additional insertion loss introduced by the switches, as well as the extra links between the nodes and the BSA. 


In this work, we focus on a distributed switching protocol for arbitrary unbuffered multidrop quantum networks. The proposed protocol handles end node entanglement requests by allocating appropriate support nodes, switches, and physical links to forward the photons from the quantum end nodes to the support nodes.
Here we consider the BSA nodes to be the main support node in the network to enable Memory-Interference-Memory (MIM) links (defined in Section ~\ref{sec:link_establishment}). The proposed protocol can also be extended for other types of link architecture such as Memory-Source-Memory (MSM) using entangled photon pair sources (EPPS) nodes.  


The remainder of the paper is organized as follows: In Section~\ref{Preliminary}, we present the prerequisites for link establishment, the problem statement, and the requirements. Section~\ref{Design} presents the proposed method for distributed switching and resource allocation in the network. Details on the implementation of the protocol and the evaluation are described in Sections~\ref{Implementation} and~\ref{Eval}, respectively. Finally, Section~\ref{Conclusion} concludes the work with future directions.

\section{Preliminaries} \label{Preliminary}


\subsection{Link Establishment}
\label{sec:link_establishment}
In order to generate entanglement between two neighbor nodes, three link architectures have been proposed: Memory-Interference-Memory (MIM), Memory-Source-Memory (MSM), and Memory-Memory (MM)~\cite{jones2016design, soon2024implementation, beukers2024remote}. Among these types of quantum link architecture, MIM is based on the process of creating local entanglement between a stationary memory and a flying qubit (here, a photon) at each node and then emitting photons toward the BSA. In general, appropriate interference of the photons at the BSA is followed by sending the classical outcome of the successful Bell state measurement results in swapping of the entanglement from the measured photons to the quantum memories at the end node~\cite{hajduvsek2023quantum}.

Although different varieties of BSA exist, implementations of this support node using linear optics and polarization encoding are usually composed of optical delay lines to control the arrival time of photons, beam splitters, and high-efficiency detectors for interference and measurement~\cite{sakuma2024optical}. 

Connecting multiple nodes requires additional considerations. A complete graph connecting $N$ nodes requires $O(N)$ network interfaces for each node and $O(N^2)$ high efficiency BSAs in total. Due to this level of complexity, creating a physical quantum link between every pair of nodes with a separate BSA is impractical. One reason is resource constraints in high efficiency detectors in cryostat. As a result, the focus of practical architectures has moved to reconfigurable interconnects using optical switches~\cite{sakuma2024optical, monga2023quant, main2025distributed}. 

From a control point of view, the establishment of a link between end nodes in a network requires two main processes. The first is the allocation of network resources, such as communication channels and BSAs, within the network. The second, often overlooked in performance evaluations, is the pre-configuration of the end nodes and BSAs for a specific connection.



\subsection{Problem Statement}

A quantum network consists of a photonic synchronization domain spread into one or multiple physical areas composed of $S$ switches and $B$ BSAs, connecting $N$ quantum nodes (end nodes, quantum repeater or quantum router \cite{vanmeter2022quantum}) in an arbitrary topology. Each pair of nodes $i,j$ has its own demands for the  generation of $\beta$ Bell pairs independent of other nodes $R_{i, j, t}$, where $t$ is the time required to access the resources (switch channels and BSAs) to create $\beta$ Bell pairs. When considering  separate scheduling for switches and BSAs, as well as the dependence of the resource usage time on candidate paths, the problem can be formulated as a resource allocation task to identify the best-fit BSAs and corresponding paths between nodes and BSAs. The network is modeled as a directed graph $G(V, E)$, where $|V| = S + B + N$ and each edge $ e_{i,j} \in E$ represents a physical channel that forwards a flying qubit from node $v_i$ to node $v_j$.  
The domain area of the problem is shown in Fig. \ref{fig:sample-switching-in-QLAN}. Finding a BSA,  reserving a path between two nodes and the BSA, agreement on the optical time-slot for the entanglement swapping among end nodes, are the main requirements to run the protocol in the network. 


\subsection{Requirements}
The specific challenges in distributed switching of entangled photons arise from three requirements:  
First, all nodes must identify available BSA resources, along with the associated cost of reaching BSAs in the network. In addition, any pair of nodes at the border of the PSD that aim to establish a link in the switched network must be aware of both nodes' ability to reach the BSAs. This implies that a strategy such as nearest-BSA-first, where 
only one node finds the closest available BSA, is insufficient. 

The second challenge is that photon routing requires allocating two distinct paths, each composed of multiple links and one destination BSA with different input ports. Since all resources such as switches and BSAs operate according to local schedules, the switches and the BSA along both paths must coordinate to ensure consistent allocation and utilization of these paths.

Finally, the specification of the request changes dynamically and directly depends on the selected paths and the BSA. The reason is that each selected path and the chosen BSA introduce different insertion losses as a cost factor, which affect the entanglement generation rate and, as a result, the time required to complete the connection, as it depends on the loss and fidelity. In the next section, a switching protocol is proposed that considers these requirements.

\section{Protocol Design} \label{Design}




To achieve distributed link establishment in the switching network, four processes are defined: BSA information sharing, shortest bi-path selection, resource allocation via bi-path reservation, and network reconfiguration followed by entanglement swapping operation. 

\begin{figure}
    \centering
    \includegraphics[width=1\linewidth]{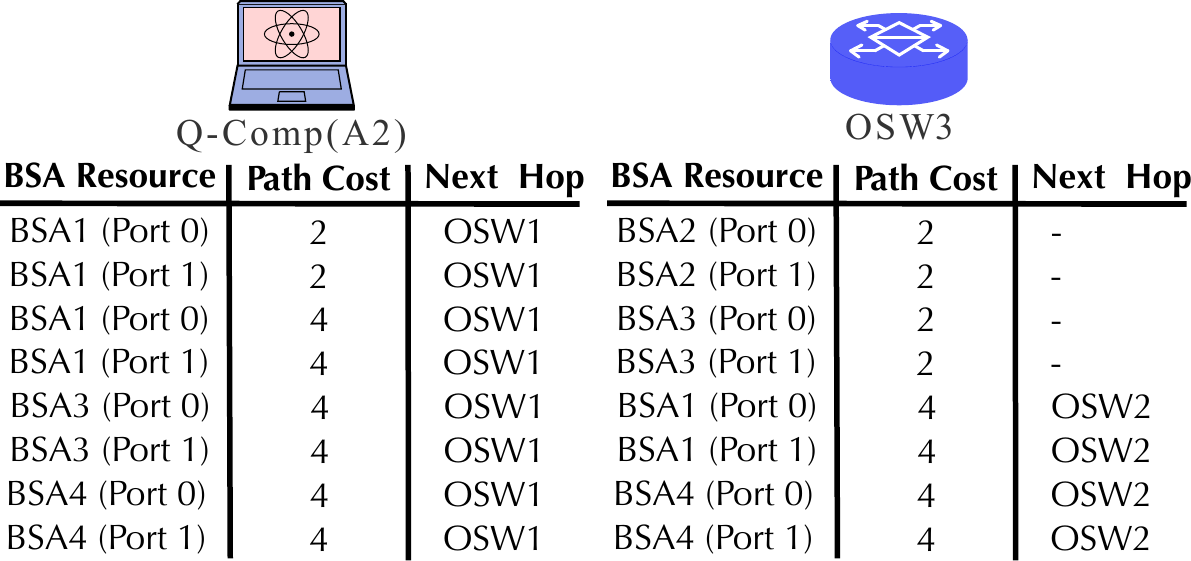}
    \caption{An example of BSA tables at an end node, Q-COMP(A2), and an optical switch (OSW3) according to the switched network in Fig. \ref{fig:sample-switching-in-QLAN}}
    \label{fig:One_BSA_Table}
\end{figure}

\subsection{Sharing BSA information} \label{ShareBSA}
In the first step, available BSAs are discovered within the network, and for each node (including end nodes at the boundary of PSD and switches), the shortest path between the node and each BSA is computed and stored in a \textit{BSA forwarding table} (in short, BSA table). An example of simplified BSA tables is shown in Fig.~\ref{fig:One_BSA_Table}. Each entry in the BSA table consists of the BSA address, physical port, the cost of the shortest path, and the next hop. This process can be initialized by running a channel discovery protocol~\cite{taherkhani2025automatic} to identify neighboring nodes connected via physical channels by considering photon propagation direction in the channel and their associated losses. The BSA table can then be constructed in a distributed manner, where each node broadcasts local connectivity information, or can be computed and installed via a software-defined networking (SDN) approach. As we show in Section~\ref{Implementation}, we follow the former approach to ensure that the BSA tables can be computed via cooperation between nodes. The communication complexity of the distributed discovery process is $O(|V|.|E|)$, due to the flooding of connectivity information across the network. After topology discovery, each node computes shortest paths to all BSA resources using Dijkstra's algorithm~\cite{dijkstra1959two} with a binary heap, resulting in a computational complexity of $O((|V|+|E|)\log |V|)$ per node. However, regardless of the approach for offline computing the BSA tables, the decisions for reserving paths and BSAs for all requests are made cooperatively between the end nodes and switches in the network, as shown in the next section.

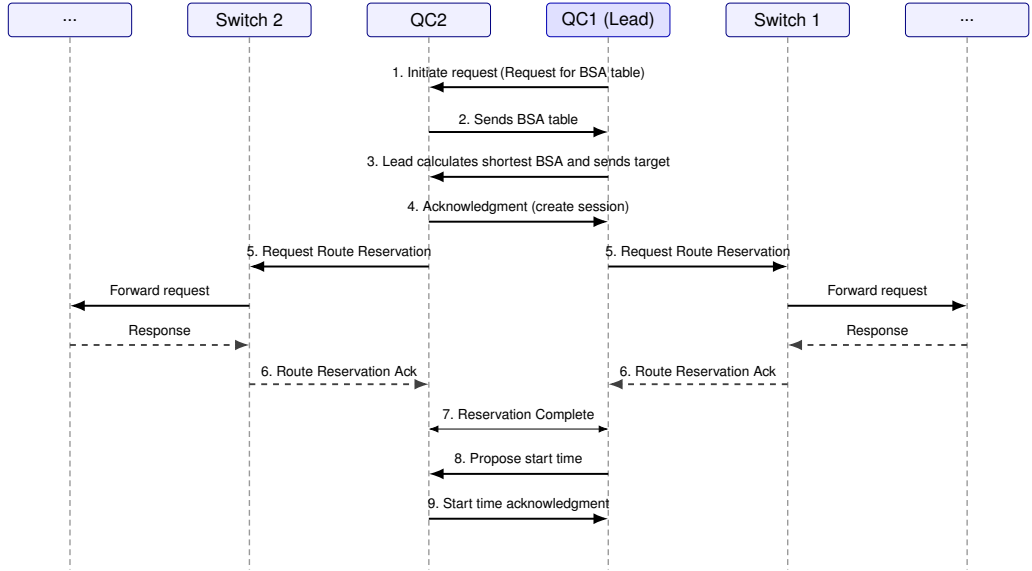
\begin{figure*}[hbt]
    \centering 
    \resizebox{0.75\textwidth}{!}{%
    \begin{tikzpicture}[
        font=\sffamily,
        >=Latex,
        participant/.style={
            draw=blue!50!black,
            rounded corners=2pt,
            fill=blue!6,
            minimum width=2.2cm,
            minimum height=6mm,
            align=center,
            font=\sffamily\small
        },
        lifeline/.style={draw=gray!80, densely dashed, line width=0.7pt},
        msg/.style={-Latex, line width=1pt, draw=black},
        resp/.style={Latex-Latex, line width=0.95pt, draw=black!75},
        note/.style={
            draw=orange!60!black,
            fill=orange!15,
            rounded corners=2pt,
            align=left,
            font=\sffamily\scriptsize,
            inner sep=4pt,
            text width=3.2cm
        }, 
    ]
    
    \node[participant] (NETL) at (0,0) {...};
    \node[participant] (S2)   at (3.2,0) {Switch 2};
    \node[participant] (QC2)  at (6.4,0) {QC2};
    \node[participant, fill=blue!12, draw=blue!70!black] (QC1) at (9.6,0) {QC1 (Lead)};
    \node[participant] (S1)   at (12.8,0) {Switch 1};
    \node[participant] (NETR) at (16.0,0) {...};
    
    \foreach \p in {NETL,S2,QC2,QC1,S1,NETR}{
        \draw[lifeline] (\p.south) -- ++(0,-9.5);
    }
    
    \def\yA{-1.2}
    \def\yB{-2.0}
    \def\yC{-2.8}
    \def\yD{-3.6}
    \def\yE{-4.4}
    \def\yF{-5.1}
    \def\yG{-5.8}
    \def\yH{-6.5}
    \def\yI{-7.3}
    \def\yJ{-8.1}
    \def\yK{-8.9}
    
    \draw[msg] ($(QC1.center)+(0,\yA)$) -- ($(QC2.center)+(0,\yA)$)
        node[midway, above, font=\sffamily\scriptsize, align=center] {1. Initiate request\,(Request for BSA table)};
    
    \draw[msg] ($(QC2.center)+(0,\yB)$) -- ($(QC1.center)+(0,\yB)$)
        node[midway, above, font=\sffamily\scriptsize] {2. Sends BSA table};
    
    \draw[msg] ($(QC1.center)+(0,\yC)$) -- ($(QC2.center)+(0,\yC)$)
        node[midway, above, font=\sffamily\scriptsize, align=center] {3. Lead calculates shortest BSA and sends target};
    
    \draw[msg] ($(QC2.center)+(0,\yD)$) -- ($(QC1.center)+(0,\yD)$)
        node[midway, above, font=\sffamily\scriptsize] {4. Acknowledgment (create session)};
    
    \draw[msg] ($(QC2.center)+(0,\yE)$) -- ($(S2.center)+(0,\yE)$)
        node[midway, above, font=\sffamily\scriptsize] {5. Request Route Reservation};
    
    \draw[msg] ($(QC1.center)+(0,\yE)$) -- ($(S1.center)+(0,\yE)$)
        node[midway, above, font=\sffamily\scriptsize] {5. Request Route Reservation};
    
    \draw[msg] ($(S2.center)+(0,\yF)$) -- ($(NETL.center)+(0,\yF)$)
        node[midway, above, font=\sffamily\scriptsize] {Forward request}; 
    
    \draw[msg] ($(S1.center)+(0,\yF)$) -- ($(NETR.center)+(0,\yF)$)
        node[midway, above, font=\sffamily\scriptsize] {Forward request};
    
    \draw[dashed, -Latex, line width=0.95pt, draw=black!75] ($(NETL.center)+(0,\yG)$) -- ($(S2.center)+(0,\yG)$)
        node[midway, above, font=\sffamily\scriptsize] {Response};
    
    \draw[dashed, -Latex, line width=0.95pt, draw=black!75] ($(NETR.center)+(0,\yG)$) -- ($(S1.center)+(0,\yG)$)
        node[midway, above, font=\sffamily\scriptsize] {Response};
    
    \draw[dashed, -Latex, line width=0.95pt, draw=black!75] ($(S2.center)+(0,\yH)$) -- ($(QC2.center)+(0,\yH)$)
        node[midway, above, font=\sffamily\scriptsize] {6. Route Reservation Ack};
    
    \draw[dashed, -Latex, line width=0.95pt, draw=black!75] ($(S1.center)+(0,\yH)$) -- ($(QC1.center)+(0,\yH)$)
        node[midway, above, font=\sffamily\scriptsize] {6. Route Reservation Ack};

    \draw[<->] ($(QC2.center)+(0,\yI)$) -- ($(QC1.center)+(0,\yI)$)
        node[midway, above, font=\sffamily\scriptsize] {7. Reservation Complete};
    
    
    \draw[msg] ($(QC1.center)+(0,\yJ)$) -- ($(QC2.center)+(0,\yJ)$)
       node[midway, above, font=\sffamily\scriptsize] {8. Propose start time};
    
    \draw[msg] ($(QC2.center)+(0,\yK)$) -- ($(QC1.center)+(0,\yK)$)
       node[midway, above, font=\sffamily\scriptsize] {9. Start time acknowledgment};

    \end{tikzpicture}
    }
    \caption{Message sequence for creating and starting a connection}
    \label{fig:communication_diagram}
\end{figure*}

\subsection{Shortest Bi-Path Selection}
This step represents the first active phase for two end nodes when they decide to establish a quantum link (in the form of a MIM link). During this phase, two nodes collaborate to select the shortest bi-path between them. They share their BSA tables and merge them into a combined BSA table. As shown in Alg.~\ref{LinkEstablishment}, the cost associated with each entry in the combined BSA table is computed as the sum of the individual costs of each end node to the BSA, representing the total cost of the two paths from both end nodes to that BSA. By completing this step, the end nodes will obtain a list of BSAs ordered by path cost, then the nodes agree on an ordered set of candidate BSAs with the shortest path costs. The BSA bi-path selection has a linear computational complexity of $O(P)$, where $P$ is the number of BSA port entries ($P \le 2B$).    

\begin{algorithm}[tb]
\DontPrintSemicolon
\SetKwInOut{Input}{Input}\SetKwInOut{Output}{Output}
    End Nodes $N_i$, $N_j$ share their BSA table \textsc{BsaT}$_i$ and \textsc{BsaT}$_j$ to each other \\
    \textsc{MBsaT}  $\leftarrow$ merge(\textsc{BsaT}$_i$, \textsc{BsaT}$_j$)\\  
    \ForEach{BSA $B_k$ in \textsc{MBsaT}} {
        $C^{\textsc{MBsaT}}_{B_k}$ = $C^{\textsc{BsaT}_{i}}_{B_k}$ + $C^{\textsc{BsaT}_{j}}_{B_k}$
    } 
    \Repeat{A logical link is established with a BSA the reserved paths}{
       \textsc{Bsa}$_{Candid}$ = pop(least cost \textsc{BSA} in \textsc{MBsaT}) 
       Send reservation request \textsc{REQ} to \textsc{BSA}$_{Candid}$ \\
       \If{\textsc{REQ} is feasible in \textsc{BSA}$_{Candid}$ schedule} {
            Find and reserve the paths between the nodes and \textsc{BSA}$_{Candid}$   
       } 
    } 
\caption{Quantum Link Establishment Algorithm (QLE)\label{LinkEstablishment}}
\end{algorithm}

\subsection{BSA and Bi-path Reservation}
Once the list of BSAs is identified, the two nodes attempt to reserve the paths that connect them to the BSA and the BSA itself. This process is similar to the reservation of two paths in a circuit-switched network.  The end nodes start reserving the path by sending reservation requests to the node defined by the next hop column of the BSA table. Each switch forwards the reservation requests until both paths reach the BSA to establish the link. 

\subsection{Network reconfiguration and operation}
After the reservation is finalized, all nodes along the paths begin to adjust their internal configurations. The calibration of the devices for the connection also occurs in this step. Once reconfiguration is complete, the paths and nodes are ready to perform entanglement swapping operation. After the entanglement swapping is performed, the result is sent to the end nodes via classical communication.

\section{Implementation} \label{Implementation}

In this section, a  detailed implementation of the protocol is described. 

\subsection{BSA Table Construction}
As mentioned in Section~\ref{ShareBSA}, here we focus on a cooperative approach to calculate the BSA tables. In this approach, each end node sends a message containing its local connectivity information to its neighboring nodes. When a node receives the message, it processes the information and forwards it to all adjacent nodes, except the incoming interface. During this step, duplicate messages are identified and dropped using a packet identification mechanism to avoid loops. At the end of this iterative process, each node accumulates global topology information from which hop counts can be calculated. Once the topology is constructed, each node then computes the shortest paths to all BSA nodes using a cost metric. In general, the cost function can be proportional to the insertion loss and can be computed during the channel discovery protocol~\cite{taherkhani2025automatic} in the physical layer. However, as shown in the next section, for simplicity, the hop count is selected as the basic metric for the cost of reaching BSAs. The resulting paths are stored in the local BSA forwarding table and later utilized for the BSA selection phase for entanglement generation. 

\subsection{Communication}

The messages sent to establish a link between two quantum end nodes and a BSA for an ideal scenario without conflicts are shown in Fig. \ref{fig:communication_diagram}.

\begin{figure}
    \centering
    \includegraphics[width=0.95\linewidth]{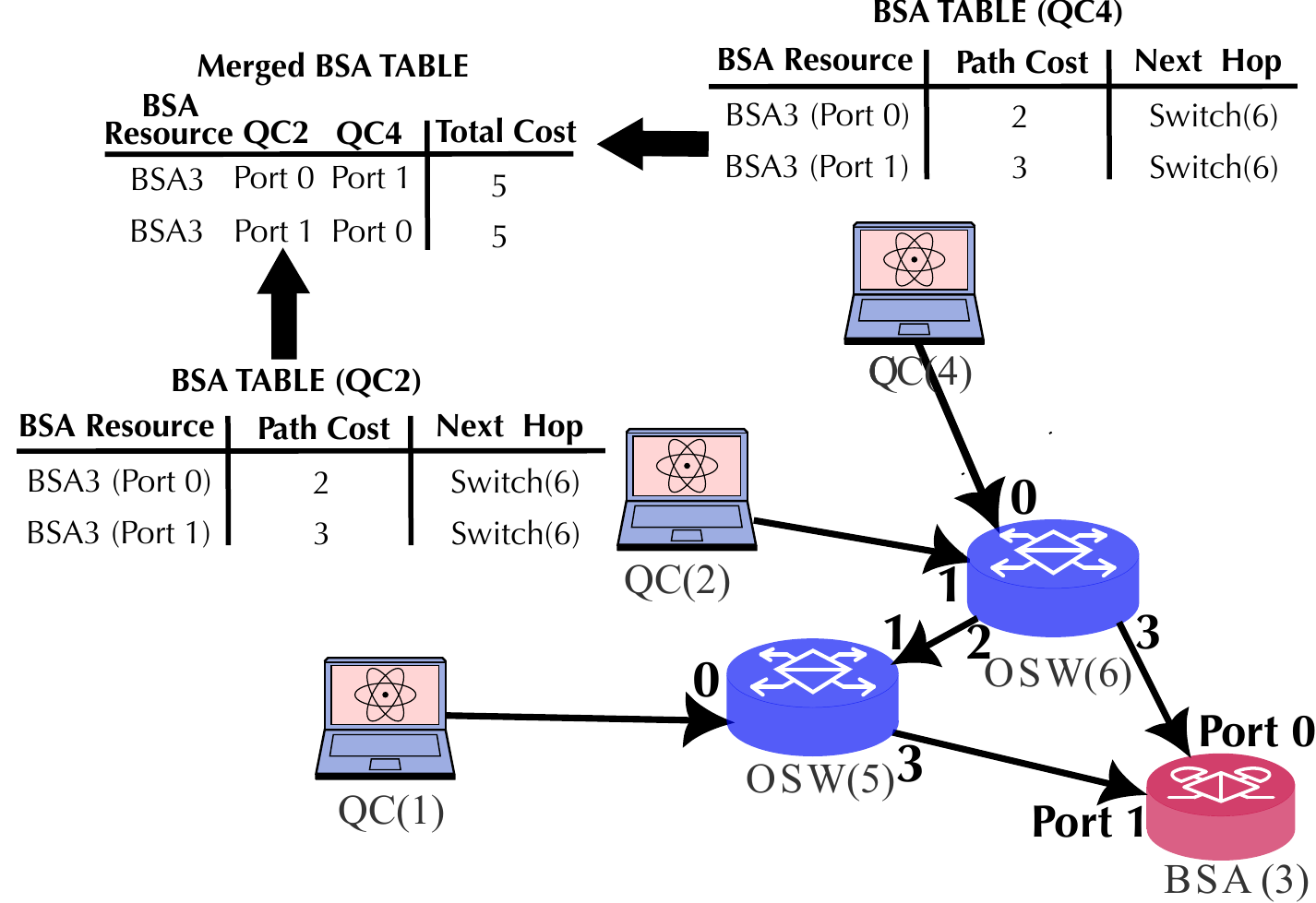}
    \caption{Simple quantum network with two switches and one BSA. 
    BSA tables of nodes QC2, QC4 are shown in the figure. Combined BSA candidate for QC(2) and QC(4) is also shown. In this case, there's only one possible candidate BSA.}
    \label{fig:simple_quantum_network}
\end{figure}

All nodes will have a BSA table at the start, before any requests can be made. The BSA table for each node contains the BSA resources that are accessible to the node. It is represented by the address of that BSA and the port number. In addition, each row contains the cost of the shortest path to reach each resource and the next hop to access the resource. Fig. \ref{fig:simple_quantum_network} shows an example of a BSA table for QC2 and QC4 in a simple network.

First, one of the two end nodes (QC2 and QC4) will assume a leading role that is responsible for starting the connection process and making several decisions during the connection setup process. Then, the lead node will send a message that requests the BSA table of the second quantum end node. If the other node is available, the BSA table will be sent back to the lead node.

The lead node will then proceed to create a merged BSA table as shown on the top left of Fig. \ref{fig:simple_quantum_network}. This is used to determine the BSA port configurations with the lowest cost considering the path from both quantum end nodes. The selected BSA and the ports to use for each quantum end node are sent to the second quantum end node. Then it will wait for an acknowledgment.

Once both end nodes have a target BSA and the target port on the BSA, they will begin to request the reservation of the route. Each end node starts by sending a route reservation request to the next hop indicated by its BSA table. Then, each of the switches that received a reservation request will pass it down until it reaches the BSA. If there are no conflicts along the way, the path is reserved, and the acknowledgment message will go through the path in reverse back to the original end node.

Finally, once the route reservation reaches the BSA and the BSA accepts the request, it will send an acknowledgment to the node that requested the reservation. This acknowledgment will be passed back the chain back to the quantum end node, indicating the success of the route reservation.

\subsection{Conflict Resolution}

As a network becomes more complex and the number of quantum end nodes making requests increases, there will inevitably be conflicts in the use of resources within the network. All three types of nodes can have conflicts. 

Firstly, for a quantum end node, one quantum end node might already be using its port for a session when a second request comes in. This means that the end node cannot establish a connection for this new request at the moment. Therefore, it will have to tell the requester to add this new request to a queue to resume at a later time.

Secondly, for a switch node, there might be multiple requests that require the use of the same switch port. This means that the switch will have to reject one of the requests. The switch will decide which request to reject on a first-come-first-served basis. The request that came in later will be rejected in favor of an earlier request.

Finally, for a BSA node, a conflict occurs when two different requests attempted to reserve ports on the BSA node. Similarly to the switch node, the BSA will also use a first-come-first-served basis and reject the later request that comes in.

Once a conflict occurred and a rejection happened, the rejection message will be relayed back to the originating quantum end node. The quantum end node will then notify its peer and finally free up all resources that are currently reserved for the rejected request. Then, the lead end node will add the request to a queue.

Once an end node becomes free, it will check inside its queue and dequeue any saved requests to continue it, as long as it is within the request's timeout period. 

\section{Evaluation} \label{Eval}

\subsection{Simulation}
We evaluate the switching protocol by developing a simulator. The tool simulates the behavior of the algorithm in a full network. It supports 3 types of nodes: quantum end nodes with quantum stationary memories such as quantum computers, optical switches, and BSAs.

The simulation places each node in a sandbox, requiring them to act independently. There is an event scheduler running on top of all nodes to send network activities from a node to another. Each node maintains only its own local state and there is no global container that contains the details of the entire network available for the nodes to use.

The nodes communicate with each other by sending or scheduling communication messages over the classical communication channel. There are several methods nodes can use to send messages to each other: by links, by the recipient's address, or broadcast to all other nodes. The simulator will pass the message to its recipient by reading the destination node address within the packets. Finally, the recipient node will handle the message inside its sandbox.

As the simulator runs, it will send certain logs to an analytics engine. Once all communications are completed, it will analyze the collected data and calculate the statistical results presented below.

\subsection{Results}
For simulation, we considered the Q-Fly topology network~\cite{sakuma2024optical}. The Q-Fly architecture is a Dragonfly-inspired scalable optical switching interconnect that organizes the network into multiple groups of quantum nodes where each group contains a group switch, shared pool of BSAs and end nodes (quantum computers)~\cite{sakuma2024optical}. We selected a single-path quasi-half duplex (SPHD) configuration with parameters $k = 6$, $N = 20$, $g = 5$, $p = 5$, and $b = 2$, and a dual-path quasi-half duplex (DPHD) configuration with $k = 12$, $N = 42$, $g = 7$, $p = 6$, and $b = 3$, where the parameters are defined in Table~\ref{tab:net_param}.  

\begin{table}[t]
\centering
\caption{Parameters for creating a sample  switched network~\cite{sakuma2024optical}}
\small
\begin{tabular}{c c}
\hline
Parameter & Definition \\
\hline
$N$ & number of end nodes in a fully populated system \\
$g$ & number of groups \\
$p$ & group size (number of quasi-half \\
    & duplex end nodes attached to one group switch)\\
$k$ & group switch radix\\
$b$ & number of BSAs in one group \\   
\hline
\end{tabular}
\label{tab:net_param}
\end{table}

With Poisson-distributed traffic sent over the network, as shown in Table.~\ref{tab:sim_summary}, we tested our protocol under three settings where $\lambda$ is 75, 100, and 150. $\lambda$, as commonly defined in the Poisson distribution, represents the average number of events per time step. However, since in our case there are multiple time steps between events, we interpret it as the average interval between requests.

\begin{figure}[tb]
    \centering
    \includegraphics[width=0.95\linewidth]{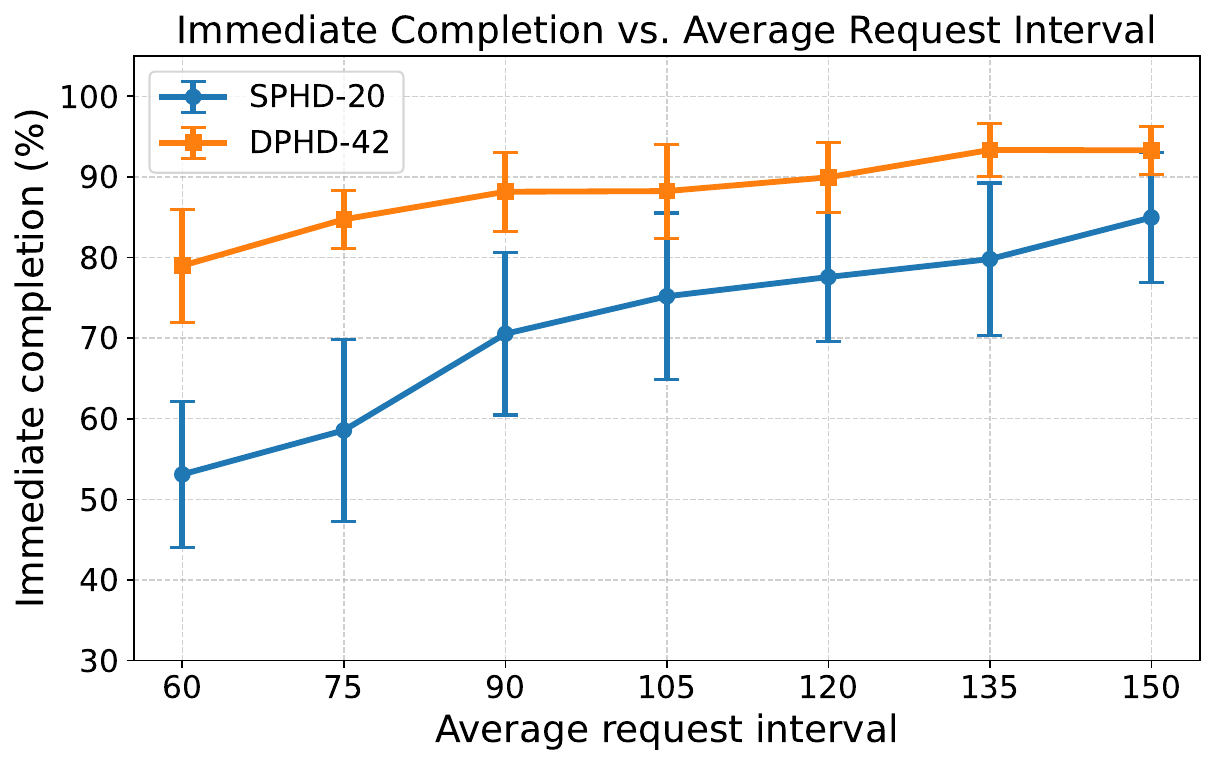}
    \caption{A performance comparison of a dense network (DPHD-42, Orange) against a sparse network (SPHD-20, Blue). The sparse network benefits from longer request interval, while the denser network exhibits stability under congested load.}
    \label{fig:effects of lambda}
\end{figure}

In all settings and in all trials, there was a 100\% success rate in generating that quantum link. 
There were no requests dropped or expired due to contention. However, several switch conflicts were observed that were handled by queueing. The maximum number of retries before success in both the SPHD-20 and DPHD-42 Q-Fly tests was two retries. The percentage of direct (immediate) completion remained high, reaching beyond the 90\% mark in some settings. Furthermore, we evaluated how varying the request interval $\lambda$ affects immediate completion in both SPHD and DPHD configurations in a wider range of traffic conditions, as illustrated in Fig.~\ref{fig:effects of lambda}.   

Queued requests were resolved using a first-come, first-served basis, where the request with the earliest arrival time is prioritized. Moreover, the average completion time for uncontested sessions was around 15 time steps for both Q-Fly settings. However, once queuing was taken into account, which includes a fixed 100 timestamp for de-queuing, the average completion time increased significantly. In addition, the drop in immediate success can be explained by resource utilization. Since the internal state of the quantum computers is not considered in the current simulation, requests are generated randomly. This, in turn, leads to heavy contention, with some BSAs handling up to 20 sessions.

\begin{table}[t]
\centering
\caption{Summary of simulation results for SPHD and DPHD configurations}
\small
\resizebox{\columnwidth}{!}{
\begin{tabular}{c c c c c c}
\hline
Topology & $\lambda$ & Requests & Immediate Completion (\%) & Comp. time & Comp. (incl. queue) \\
\hline
SPHD-20 & 75  & 120--133 & 85--89 & $<15$ & 125--135 \\
DPHD-42 & 75  & 118--130 & 60--74 & $<15$ & 150--170 \\
SPHD-20 & 100 & 91--104  & 85--94 & $<15$ & 125--135 \\
DPHD-42 & 100 & 91--98   & 73--79 & $<15$ & 140--150 \\
SPHD-20 & 150 & 53--70   & 85--88 & $<15$ & 122--130 \\
DPHD-42 & 150 & 62--79   & $>90$  & $<15$ & 125--175 \\
\hline
\end{tabular}
}
\label{tab:sim_summary}
\end{table}

\section{Conclusion and Future Works} \label{Conclusion}
In this work, we addressed the problem of quantum link establishment in a multidrop, memoryless switched network. The proposed protocol allocates Bell state analyzers (BSAs) and switch links as shared resources in a distributed manner. This is achieved through a process consisting of BSA table construction, shortest bi-path selection, and reservation of the paths. As a result, the network provides a time-division multiplexing of circuit switched resources. For future works, handling of requests based on different queuing mechanisms should be analyzed in more detail.  

\section*{Acknowledgment}
The authors would like to thank Andrew Todd, Daisuke Sakuma, Hiroyuki Ohno, Yuki Kurosawa and Kentaro Teramoto for the fruitful discussions, and Kent Oonishi, Saori Sato, Kaori Nogata, and Kaori Sugihara for the management and support.

\bibliographystyle{IEEEtran-new.bst}
\bibliography{IEEEabrv, bibfile}

\end{document}